\documentclass{iau}
\usepackage{graphicx}
\usepackage{amsmath}
\usepackage{subcaption}
\usepackage{multirow}
\usepackage{csquotes}

\begin{document}

\lefttitle{Kova\v{c}evi\'c, Mason, {\'C}iprijanovi{\'c}, Long, Korczakowska, Moore, Vuolukka}
\righttitle{Astrobiology and Technosignatures in the Rubin LSST  }

\jnlPage{1}{7}
\jnlDoiYr{2026}
\doival{10.1017/xxxxx}

\aopheadtitle{Proceedings IAU Symposium No.\,404}
\editors{J.\ Haqq-Misra \& R.\ Kopparapu, eds.}

\title{Prospects for Astrobiology and Technosignature Searches with the Vera C. Rubin Observatory Legacy Survey of Space and Time}

\author{Andjelka B.\ Kova{\v c}evi{\'c}$^1$,
        Nigel J.\ Mason$^2$,
        Aleksandra {\'C}iprijanovi{\'c}$^{3,4,5}$,
        Becky Long$^2$,
        Dominika Korczakowska$^2$,
        Maia Moore$^2$,
        Juulia Vuolukka$^2$}

\affiliation{%
  $^1$University of Belgrade -- Faculty of Mathematics, Department of
       Astronomy, Belgrade, Serbia;
       email: {\tt andjelka.kovacevic@matf.bg.ac.rs}\\
  $^2$School of Physics and Astronomy, University of Kent,
       Canterbury, CT2\,7NH, United Kingdom;
       email: {\tt N.J.Mason@kent.ac.uk}\\
  $^3$Fermi National Accelerator Laboratory, Batavia, IL\,60510, USA\\
  $^4$Department of Astronomy and Astrophysics, University of Chicago,
       Chicago, IL\,60637, USA\\
  $^5$NSF-Simons AI Institute for the Sky (SkAI),
       172 E.\ Chestnut St., Chicago, IL\,60611, USA}

\begin{abstract}
The Vera C.\ Rubin Observatory Legacy Survey of Space and Time (LSST) will repeatedly measure the brightness of billions of astrophysical sources using six colour filters. {We test a simple idea: real physical signals tend to appear as coordinated patterns across several measurements, whereas noise does not; we call this behaviour ``coherence''. We illustrate the approach with three LSST-like simulations: (i) five Kuiper Belt Object (KBO) surface/activity states, (ii) 648 exoplanet spectra with an injected vegetation-red-edge-like spectral feature, and (iii) 500 multiband light curves. In idealised simulations, a weakly active KBO is strongly separated from its inactive counterpart when all five colours are combined. The injected red-edge-like feature becomes detectable at effective visible-disc coverage of about 13\%, and stacking about 150 detections could reach 5$\sigma$. Coherent, natural, and noise-like variability occupy distinct regions. A small Gaia DR3 stellar sample illustrates how colour and stability may help prioritise Galactic search regions.}
\end{abstract}


\maketitle

\section{Introduction}
\label{sec:intro}

The Vera C.\ Rubin Observatory Legacy Survey of Space and Time
\citep[LSST,][]{ivezic2019} will repeatedly photograph the sky {using six}
  filters ($u,g,r,i,z,y$) over roughly ten years,
building up brightness measurements and light curves for an
unprecedented number of astrophysical sources.
This creates an opportunity for astrobiology and technosignature
searches --- the search for signs of life or of alien technology ---
but it also raises a practical question: with billions of objects,
how do we decide which ones are worth a closer look?

The traditional approach flags an object as a candidate if a single
measurement looks unusual: an oddly red colour, an oddly bright epoch,
and so on. The approach we test here instead looks for
\emph{coherence}: whether several independent measurements of the
same object move together in a way that is hard to explain by chance
or by ordinary astrophysical processes. For example, a real physical
change on a planet or asteroid's surface does not just shift one
colour at random; it tends to shift several colours together, in a
consistent direction, because it is driven by a single underlying
cause (a change in surface composition, a cloud of dust, an
absorbing gas, and so on). This means the ``fingerprint'' of a
genuine signal is not a single number, but a consistent
\emph{direction} of change across multiple observables.  {In this paper we build upon}
this idea, following related recent work \citep{2025Gallay,kovacevic2025,li2022}.

This logic is especially useful for LSST because each of its colour
filters is not a sharp, single-wavelength measurement: it is a broad
band that adds up light over a wide range of wavelengths. This means a lot of fine spectral detail {is 'washed out'}.
{However, broad physical changes --- such as a change in how red or blue
a surface or atmosphere is overall, the presence of a hazy cloud, an
absorbing gas, or a sharp red-edge-like reflectance feature --- can
still leave a detectable, correlated imprint across several LSST filters
at once, even if no single filter captures the full picture
\citep{seager2005}}. LSST photometry therefore carries
information not just in how bright or how red an object is, but in
how its colours move together, or ``cohere'' --- a logic that has
also been explored for reflected-light technosignatures {in} future
direct-imaging missions \citep{haqqmisra2022,kopparapu2024}.
While this  paper focuses on astrobiology-motivated
examples, the same coherence logic applies to technosignature
scenarios, such as unusual industrial atmospheric chemistry or
a partial dimming pattern caused by a large artificial structure
(a so-called Dyson-sphere-like scenario): both would also produce a
correlated, directional shift in colour or brightness-over-time
measurements of the kind we {discuss} here.

We illustrate {these ideas with} three prototype (proof-of-concept)
test cases, described in Section~\ref{sec:framework}: {(i) a set} of
simulated Kuiper Belt Object surface changes, {(ii) a set} of simulated
exoplanet spectra with an injected vegetation red edge (VRE)-like spectral feature,
and {(iii) a set} of simulated multiband light curves. In each case {we explore
whether} looking at the \emph{combined, coordinated} pattern across
colours or bands gives a more reliable diagnostic for Rubin-era
astrobiology and technosignature searches than looking at any single
measurement.

\section{Framework and prototype workflows}
\label{sec:framework}

{Before discussing the technical details  let us summarise the idea.} For
each of our three test cases, we simulate what LSST would measure
for an object in two states: a ``normal'' state and a state with
something added (activity, an atmospheric feature, a variability
pattern). We then ask two questions: (1)~\emph{How big} is the
combined shift in the LSST measurements between the two states,
relative to the noise? and (2)~\emph{Does the shift move in a
direction} that  ordinary astrophysical effects (like haze or
methane, in the exoplanet case) could not easily produce? A shift
that is both large and points in a distinctive direction is what we
call ``coherent'', and is more convincing than a single odd number in
one filter. The rest of this section turns that idea into three
concrete recipes, one per test case; readers who only want the
results can skip to Section~\ref{sec:results} and refer back to the
equations here as needed.

\subsection{Coherence diagnostics}
\label{sec:diagnostics}

{Testing several correlated measurements together is almost always more powerful than testing each one alone.} {Instead of flagging a
candidate object as interesting simply because one single measurement looks unusual, we test a simple idea: real, physically-caused signals tend to show up as a coordinated pattern across several measurements at once (for example, several colours shifting together in a consistent direction, or several photometric bands varying together over time), whereas noise does not}. {We use this logic {in cases discussed below},  {so that measurements are reduced to a single number} that reflects how ``coherent'' the combined set
of measurements is}. {In each of}  our three test cases, we give
the essential idea and the one number it produces; the full
statistical derivations --- including how we account for correlated
noise between bands and for ``look-alike'' effects such as haze or
methane --- are worked out in detail in a companion paper
\citep{long2026}, and are only summarised here.

Throughout, we assume a single-epoch, per-band photometric
uncertainty of $\sigma_m=0.01$\,mag, typical of LSST's design
goals. Combining bands into colours, and differencing two independently measured
colours, increases this noise slightly at each step (to $\simeq0.014$ mag
and $0.02$ mag respectively) through standard error propagation.

\begin{figure*}[ht]
\centering
\includegraphics[width=0.95\linewidth]{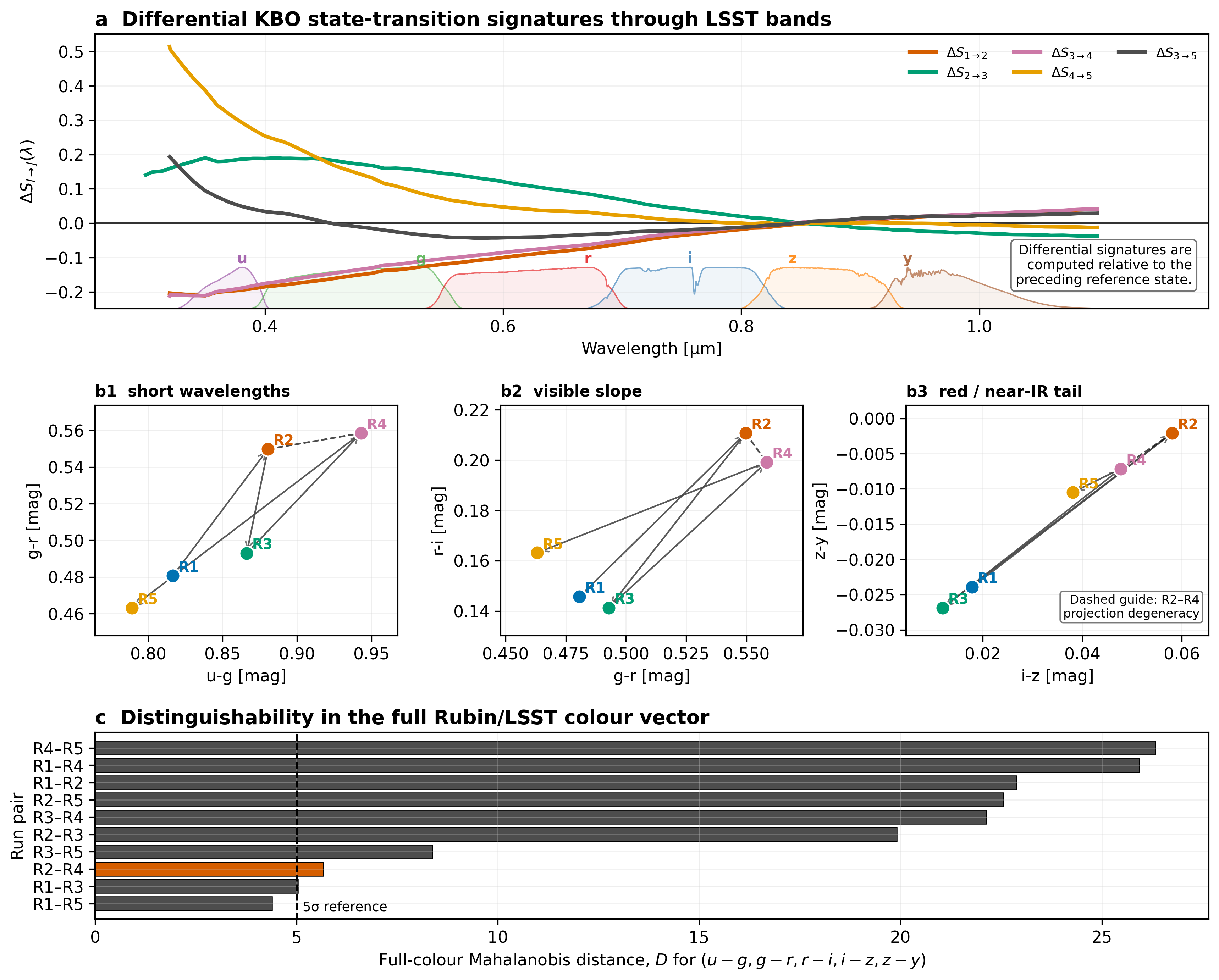}
\caption{KBO  simulation.
(a)~Differential state-transition signatures from forward PSG
reflected-light spectra projected through Rubin/LSST total system
throughput curves.
(b1)--(b3)~Simulated states in complementary LSST colour--colour
spaces; the visible-slope plane gives the clearest separation,
while redder spaces show increasing compression and degeneracy.
(c)~State distinguishability in the full LSST colour vector via
Mahalanobis distance ($D\equiv{\rm SNR}_{\rm coma}$,
Eq.~\ref{eq:coma_snr}), with $\sigma_m=0.01$ mag per band. The dashed line marks the distance corresponding to a one-sided $5\sigma$ Gaussian tail probability for $p=5$ colours, $D\approx6.2$. }
\label{fig:kbo_coherence}
\end{figure*}

\textit{Kuiper Belt Object (KBO) experiment (Eq.~\ref{eq:coma_snr}).}
Here we apply the joint-testing logic above directly: instead of
comparing each of the five LSST colours to its own noise one at a
time, we combine all five colour shifts into a single joint test
statistic, properly weighted by their (co)variance. This combined
score is the quantity $D$ in Eq.~\ref{eq:coma_snr}, the Mahalanobis
distance, which can be read simply as ``how many combined standard
deviations apart the two states are'', now using all five colours
together instead of just one:
\begin{equation}
D={\rm SNR}_{\rm coma}=\sqrt{\Delta\mathbf{c}_{\rm coma}^{T}\Sigma^{-1}
\Delta\mathbf{c}_{\rm coma}},
\label{eq:coma_snr}
\end{equation}
where $\Delta\mathbf{c}_{\rm coma}$ collects the colour shifts
between the two states and $\Sigma$ is the noise budget described
above. {For example, with $p=5$ colours combined, a genuine $5\sigma$ Gaussian tail
probability corresponds to $D\approx6.2$, because in $p$ dimensions
the reference $\chi^2$ distribution has $p$ degrees of freedom. This differs
from the one-dimensional VRE case below, where $5\sigma$ corresponds simply to
$\mathrm{SNR}_{\rm VRE}=5$. The larger KBO value reflects the dimensionality of
the colour vector. Here the reference state is treated as a deterministic forward-model prediction.
If equal independent photometric uncertainties are also assigned to the
reference state, the covariance of the colour difference doubles and the
resulting Mahalanobis distances decrease by a factor of $\sqrt{2}$.}

\begin{figure*}[ht]
\centering
\includegraphics[width=\linewidth]{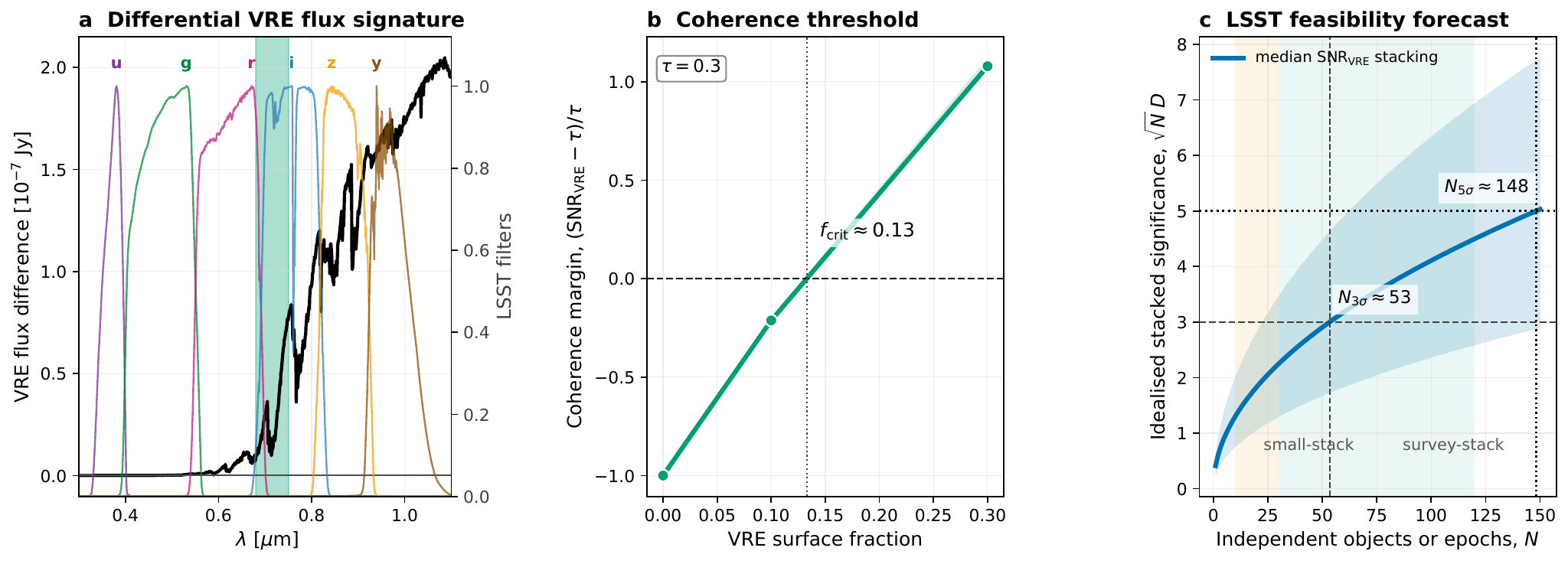}
\caption{VRE-like test-feature simulation, built from PSG model
atmosphere spectra and then converted into Rubin/LSST broadband
colours (see Section~\ref{sec:simsetup} for the full pipeline).
(a)~{The black curve is the difference between the PSG model
spectrum with and without the injected VRE-like
spectral feature, at native PSG resolution $R=1000$ before
any broadband integration}  for a representative
0.1\,bar, 95.5\% H$_2$/He, methane-poor, haze-free model with a
small illustrative injected-feature amplitude ($f_{\rm VRE}=0.02$). The coloured curves are the LSST $u,g,r,i,z,y$ filter
throughputs. 
(b)~VRE colour-coherence margin $M_{\rm VRE}$ versus fractional
VRE-like coverage $f_{\rm VRE}$ (Section~\ref{sec:simsetup}). For
the illustrative threshold $\tau=0.3$, the median score crosses the
threshold at $f_{\rm crit}\approx0.13$.
(c)~Idealised forecast  of how the combined
significance $\sqrt{N}\,D_0$ (built from $\mathrm{SNR}_{\rm VRE}$,
Eq.~\ref{eq:vre_score}) would grow with the number $N$ of
independent objects or epochs combined. Blue band is 16--84 percentile
range across the simulated grid. Pale vertical bands are illustrative
small-stack and survey-stack regimes.}
\label{fig:vre_coherence}
\end{figure*}

\textit{Vegetation red edge-like (VRE) experiment (Eq.~\ref{eq:vre_score}).}
Here the complication is that our signal of interest (the VRE-like
feature) is not the only thing that can shift the relevant colours:
haze and methane can produce a similar-looking shift. This is
similar to tuning a radio receiver to pick out one station's signal
while suppressing static from nearby frequencies: we build a
specific combination of two colour differences,
$\Delta(r-i)$ and $\Delta(i-z)$ (Section~\ref{sec:simsetup}), chosen
so that it responds strongly to the VRE-like feature but as weakly
as possible to haze and methane. This ``tuned'' signal, $S_{\rm
VRE}$, and its significance are
\begin{equation}
S_{\rm VRE}=W_1\,\Delta(r-i)+W_2\,\Delta(i-z),
\qquad
{\rm SNR}_{\rm VRE}=\frac{S_{\rm VRE}}{\sigma_{S_{\rm VRE}}},
\label{eq:vre_score}
\end{equation}
where $W_1$ and $W_2$ set how much weight $\Delta(r-i)$ and
$\Delta(i-z)$ receive in the combined signal $S_{\rm VRE}$, and
$\sigma_{S_{\rm VRE}}$ is the uncertainty on that combination. The
weights are determined from the simulated grid so that the score
responds strongly to the injected VRE-like feature while reducing
its response to haze and methane, which can  produce part
of the same colour change. Thus, $S_{\rm VRE}$ is a targeted
broadband-colour diagnostic, rather than a resolved spectral
measurement. From ${\rm SNR}_{\rm VRE}$ we
define a coherence margin using an illustrative threshold
($\tau=0.3$, chosen as the point where the median VRE-like signal
starts to separate from the baseline in our simulations). We then take $D_0$, the typical (median)
per-object ${\rm SNR}_{\rm VRE}$ across the simulated grid, as a
per-object coherence score, and forecast how the combined
significance would grow as $\sqrt{N}\,D_0$ if $N$ independent
detections were stacked together, under the optimistic assumption
that no additional systematic error floor is present.

\textit{Light-curve experiment}. {Here the analogy is several radio stations that either
broadcast the same programme in sync, or each play their own independent noise:
we ask whether an object's brightness  in different LSST bands {rises and
falls together} (``in sync'') or independently. We summarised each simulated light
curve using two quantities:   
$C_{\rm chroma}$, which measures how much its colour
changes with time, and $C_{\rm cross}$, which measures how closely its brightness
variations are synchronised across bands.}

For the quantitative diagnostic, we form near-simultaneous colour pairs,
accepting measurements separated by at most $\Delta t_{\rm pair}=0.35$~d.
The chromatic-variability coordinate is
$
C_{\rm chroma}=\sigma_{g-r}+\sigma_{r-i},
$
where $\sigma_{g-r}$ and $\sigma_{r-i}$ are the rms scatters of the paired
$g-r$ and $r-i$ colour series. Thus, low $C_{\rm chroma}$ indicates nearly
achromatic variability, whereas high values indicate stronger colour changes.

The cross-band coherence  is
$
C_{\rm cross}=\frac{1}{2}\left({\rm XCI}+{\rm SCI}\right),
$
where ${\rm XCI}$ (the cross-band coherence index) is a zero-lag discrete
correlation function  between 
 the r-band magnitude series and the g- and  i -band magnitude
series
\footnote{The discrete correlation function measures how closely two
irregularly sampled magnitude series pairs --- here r with g, or
r with i --- rise and fall together at a specified time lag.}, and
${\rm SCI}$ (the same-night coherence index) is the Pearson correlation of
strictly near-simultaneous pairs of r--g and r--i magnitude pairs, matched across bands within
$\Delta t_{\rm pair}=0.35$~d. Averaging the two combines a lag-tolerant
correlation (${\rm XCI}$, which tolerates cadence offsets up to a day) with a
strictly simultaneous one (${\rm SCI}$, which cannot be inflated by
interpolation), so a high $C_{\rm cross}$ requires genuine cross-band
synchrony rather than a cadence artefact.
Values of $C_{\rm cross}$ close to $+1$ indicate strongly synchronised
multiband variability, while values near zero indicate weak or incoherent
cross-band behaviour.
\subsection{Simulation setup}
\label{sec:simsetup}

Before describing each case, {we can describe the overall pipeline for the KBO and VRE
experiments as it is the same for both}. 
For each physical state we first generate a high-resolution
\emph{model spectrum} (at spectral
resolution $R=1000$, i.e.\ far finer than any broadband filter)
using the Planetary Spectrum Generator \citep[PSG,][]{villanueva2022}
radiative-transfer code. Each spectrum is then
passed through the LSST filter throughput curves (one curve per
band: $u,g,r,i,z,y$) and integrated, which is the standard way of
converting a model spectrum into the single number that a real
broadband observation in that filter would measure. This produces
one synthetic magnitude per band, and from these we form colours
(differences between adjacent bands, e.g.\ $g-r$). This is the only
photometric quantity actually used in the statistical analysis of
Section~\ref{sec:diagnostics}. 

\textit{KBO.}
Following this pipeline, PSG reflected-light spectra were generated
for an idealised KBO at 40\,AU and then integrated through the full
Rubin/LSST $ugrizy$ throughputs (filter, optics, detector, and
atmosphere at 1.2\,airmasses) to obtain synthetic $ugrizy$
photometry. Spectra for five surface/activity states were modelled,
intended to span a plausible range of KBO surface compositions and
a possible outgassing event: R1, a 100\% achondrite (rocky,
meteorite-like) inactive baseline; R2, a mixture of 40\% organic
material and 60\% achondrite; R3, a mixture of 30\% Antarctic-ice
analogue and 70\% achondrite; R4, a three-component mixture of 40\%
achondrite, 30\% ice analogue, and 30\% organic material; and R5,
the same mixture as R4 but with a weak coma added (a thin cloud of
silicate dust around the object, mimicking a mild outgassing or
activity event).

\textit{VRE.}
For the vegetation-red-edge-like case, we used a PSG template
modelled on the sub-Neptune exoplanet GJ\,1214b purely as a
convenient, self-consistent atmosphere to inject our test feature
into --- not as a claim that GJ\,1214b itself hosts vegetation.
Model spectra were generated over $0.30$--$1.10\,\mu$m at $R=1000$.
Thick-atmosphere sub-Neptunes of this kind can maintain a temperate
pressure layer (roughly 0.1--1\,bar) where liquid water could in
principle exist, which is why such planets have been discussed
speculatively as possible hosts for cloud-dwelling, pigmented
organisms \citep{seager2021}. 
We stress that the S-shaped feature is simply an injected VRE-like
spectral feature of known shape, centred at the wavelength where
Earth's vegetation red edge occurs
$(\lambda_0=0.705\,\mu{\rm m})$, which also happens to sit at
the boundary between the $r$ and $i$ filters, maximising its effect
on the $r-i$ colour.

{
To build the test grid, we first varied four atmospheric parameters
without an injected VRE-like feature: pressure (0.1, 1, and 10\,bar),
H$_2$/He composition (95:5, 80:20, and 60:40), methane abundance
(PSG input scales $10^{-4} $to $10^{-1}$, four values), and haze optical depth
(0.0, 0.3, and 1.0). This gives
$3\times3\times4\times3=108$ baseline spectra. We then injected a
smooth S-shaped VRE-like spectral feature into each baseline spectrum,
varying its amplitude,
$A\in\{0.0,0.10,0.30\}$ 
and transition width
$w\in\{0.02,0.05\}\,\mu{\rm m}$. This produces 648 grid entries
$(108\times3\times2)$, of which 540 are unique because the width is
irrelevant when $A=0$.
The observed amplitude of a disc-integrated red-edge signal depends on
both the intrinsic contrast of the contributing component and the
fraction of the visible, illuminated disc over which it contributes,
with additional dilution from clouds and other surface components
\citep{seager2005}. In this idealised experiment, we neglect such
additional spatial mixing and fix the intrinsic contrast of a fully contributing
component to unity. Accordingly, the injected amplitude $(A)$ is set
equal to $(f_{\rm VRE})$, which we interpret as an effective
visible-disc coverage parameter. Thus, the grid samples
$f_{\rm VRE}=A\in\{0.0,0.1,0.3\}$, where $f_{\rm VRE}=0$ means that no
VRE-like feature is injected, while larger values give a larger
disc-integrated red-edge-like contribution.}

\textit{Light curves.}
For the third case, we simulated 500 LSST-like $g,r,i$ light curves
directly (as brightness-versus-time sequences, without an underlying
spectrum), split evenly into three classes of $\approx$167 objects
each, over a 3-year LSST-like observing baseline with irregular
cadence (mean 1.2 visits per night, with seasonal gaps of 185 days
when a given field is not observable) and typical LSST band
fractions ($g$:$r$:$i$ = 15\%:55\%:30\%). \emph{Coherent} objects
follow a stable periodic brightness variation with semi-amplitude
$A=0.10$\,mag (plus a smaller 0.03\,mag harmonic) at a period drawn
uniformly from 30--90 days, with the \emph{same} variation shape and
size applied identically in all three bands (i.e.\ achromatic ---
the object gets brighter and fainter together in $g$, $r$, and $i$).
\emph{Natural} objects  follow a periodic signal whose
amplitude and phase drift randomly over time, on top of which we add
a chromatic (colour-dependent) modulation meant to mimic
cloud/weather-like effects, scaled differently in each band
($s_g=1.40$, $s_r=1.00$, $s_i=0.65$); this produces a
brightness--colour coupling (the object's colour changes as it gets
brighter or fainter) that is absent by construction in the coherent
class. \emph{Noise} objects have constant underlying brightness,
with only photometric measurement scatter (signal-to-noise ratio
between 10 and 50). For every simulated object we then computed the
summary statistics defined in Section~\ref{sec:diagnostics}
($C_{\rm chroma}$, $C_{\rm cross}$), together with several
supporting diagnostics --- an irregular-cadence autocorrelation, a
phase-stability index, an amplitude-stability index, the paired
colour scatter, and a brightness--colour coupling coefficient --- and
placed each object in the $C_{\rm chroma}$--$C_{\rm cross}$ plane
used in Fig.~\ref{fig:lightcurves_coherence}.

\section{Results and discussion}
\label{sec:results}

Figure~\ref{fig:kbo_coherence} shows that as the KBO's simulated
surface/activity state changes step by step (R1$\to$R5), its LSST
colours do not jump around {randomly, rather they trace out} a smooth,
low-dimensional track through colour space, i.e. the different
states line up along a simple, predictable path rather than
scattering unpredictably. The panel showing $g-r$ versus $r-i$ (the
``visible-slope'' colours) gives the clearest separation between
states, while colours further to the red become increasingly bunched
together and harder to tell apart. {The weak-coma transition from R4 to R5 --- i.e.\ the same surface composition
with and without the added coma --- produces the strongest separation under
the simulated conditions, with a full-colour Mahalanobis distance of
$D \approx 26$ relative to the noiseless R4 forward-model reference. If
equal observational uncertainty is assigned to both states, this value becomes
$D \approx 18.6$. Several other state pairs are also strongly distinguishable, whereas the least separated pairs fall below the $5\sigma$ reference ($D\approx6.2$). Notably, R1--R5 is among the least separable pairs: R5's organic
fraction is modest and diluted within the ice--achondrite mixture,
and its silicate coma shifts the broadband colours back toward the
rocky R1 baseline, leaving offsets of only ${\sim}1\sigma$ per
colour. Still, broadband coherence flags that a photometric state
change has occurred, rather than uniquely identifying which.} Thus, in this simulation, weak KBO activity produces a pronounced coordinated displacement in the full LSST colour vector rather than an isolated anomaly in a single colour.

Figure~\ref{fig:vre_coherence} applies the same logic to the
VRE-like test feature. Panel (a) shows a high-resolution model
difference spectrum (the black curve) only to locate the injected
feature relative to the LSST passbands; it is not a simulated LSST
observation. Once this model is integrated through the broad LSST
filters, it produces correlated shifts across several broadband
measurements rather than a feature resolved in any single band.
Panel (b) shows the VRE coherence margin
$M_{\rm VRE}=({\rm SNR}_{\rm VRE}-\tau)/\tau$ as a function of the
fractional VRE-like coverage $f_{\rm VRE}$ (defined in
Section~\ref{sec:simsetup}). For the adopted illustrative threshold
$\tau=0.3$, the median VRE-sensitive score crosses the threshold at
$f_{\rm crit}\approx0.13$,
an effective visible-disc coverage of approximately $13$\%
under the unit-contrast assumption. This value is an
idealised interpolation between the simulated grid points at
$f_{\rm VRE}=0.1$ and $0.3$, rather than a calibrated general
detectability limit. Panel (c) shows how the achievable significance would grow if many such
detections were combined: i.e. no residual calibration or modelling uncertainty
that would remain after stacking, the 3$\sigma$ and 5$\sigma$ levels are
reached after combining roughly $N \approx 53$ and $N \approx 148$
independent objects or epochs, respectively. The shaded band reflects how much this estimate varies
across our simulated atmospheric grid.

\begin{figure*}[ht]
\centering
\includegraphics[width=\linewidth]{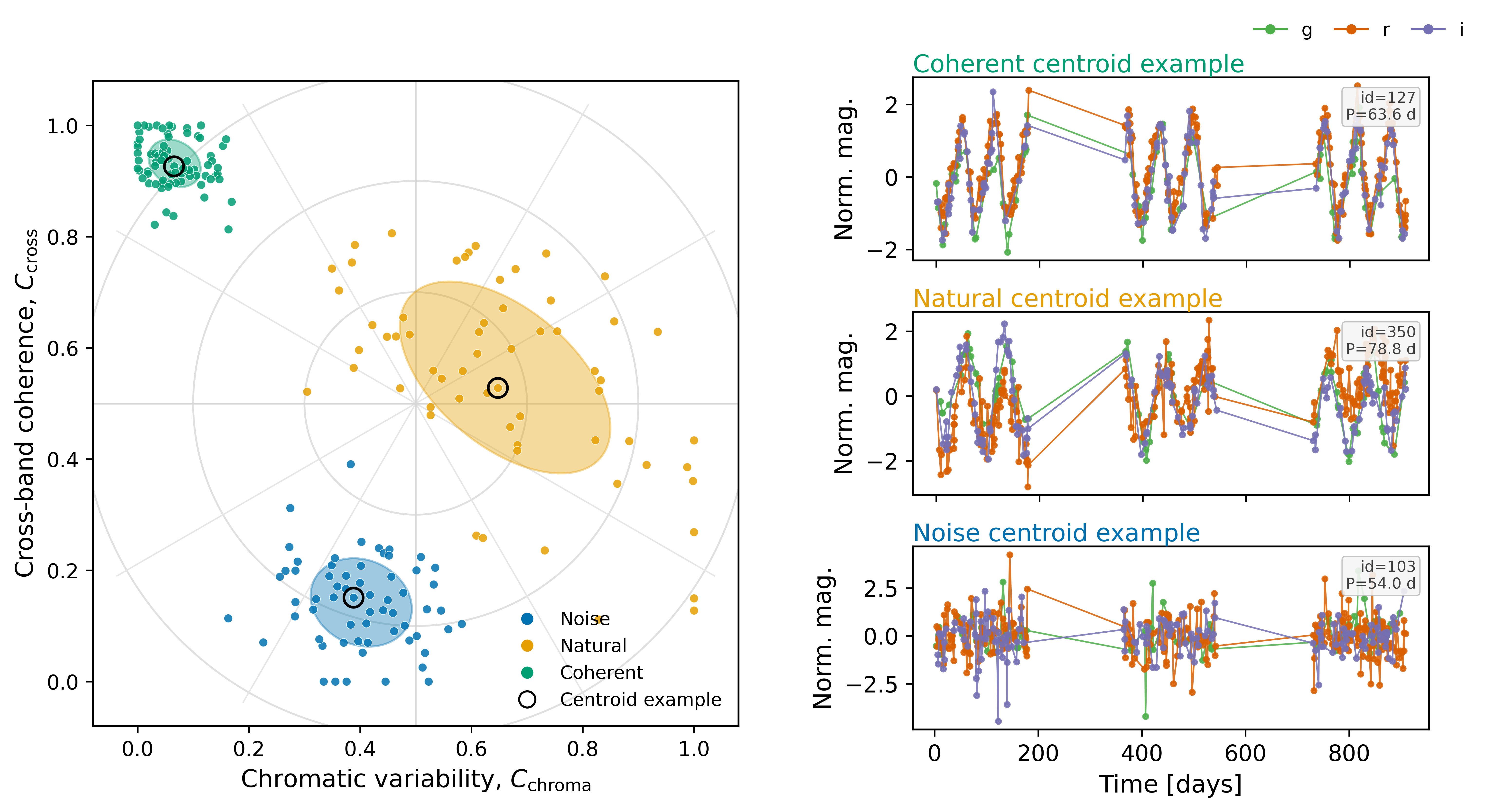}
\caption{Simulated multiband light-curve populations in the
$C_{\rm chroma}$--$C_{\rm cross}$ diagnostic plane. {For visual comparison, both diagnostic coordinates are robustly
rescaled to [0,1] using the 2nd and 98th percentiles of the
simulated sample.}
\textit{Left}: noise-dominated, natural, and coherent variability classes  coloured by true (injected) class
($N\approx167$ each, 500 total). Ellipses show population covariance,
open circles mark centroid-nearest examples.
\textit{Right}: normalised Rubin/LSST $g$, $r$, $i$ light curves for those
examples. }
\label{fig:lightcurves_coherence}
\end{figure*}

{Figure~\ref{fig:lightcurves_coherence} extends the same idea to
brightness variation over time, comparing three simulated classes:
a stable, coordinated (coherent) periodic signal, natural
stochastic (randomly drifting) variability, and pure noise. The same
approach could later be applied to hypothetical engineered brightness
patterns --- not to claim a detection of such a pattern, but
to test whether different physical or artificial mechanisms occupy
distinguishably different regions of this diagnostic space.
Individual measurements such as phase stability or overall variance
overlap a lot between these three classes and do not, on their own,
reliably tell them apart. However, when we look at $C_{\rm chroma}$
and $C_{\rm cross}$ \emph{together} --- i.e.\ how much an object's
colour changes over time, plotted against how well its brightness
changes track each other across bands --- the three classes separate
into distinct regions: coherent objects {maintain a well-aligned behaviour}
across bands, natural objects show structured but less stable
behaviour, and noise-dominated objects show no persistent
cross-band pattern. {However, our three} simulated classes do not cover
the whole diagnostic plane. Many ordinary periodic variables,
such as pulsating stars and spotted rotating stars, vary strongly in
both brightness and colour, with the bands well synchronised. These
objects would fall in a fourth region of the plane --- high
$C_{\rm chroma}$ and high $C_{\rm cross}$ --- and are the main
false-positive population that the tests planned in
Section~\ref{sec:conclusion} must address.}

\subsection*{{Where}
in the Galaxy to look for anomalies}
\label{sec:hpi}

The three coherence experiments above establish \textit{how} to
detect anomalous signals, {however} a complementary question is \textit{where}
in the Galaxy to look.

{As a proof-of-concept, we use a small Gaia DR3 \citep{maiz2023}
stellar sample (Table 
~\ref{tab:gaia_habitability_context}).
We ask whether a Galactic field contains stars with two
separately favourable properties: solar-like colour and low photometric
variability.
We therefore summarise each Galactic-latitude bin by two fractions related to those two properties. The first is
$f_{\rm solar}$ describing the fraction of stars with Gaia colour
$0.8 < BP-RP < 1.5$ \citep{mucciarelli2021}, corresponding approximately to
G-to-mid-K main-sequence colours for low-reddening stars. The second is  $f_{\rm calm}$ describing  the fraction of stars
with low Gaia $G$-band variability, defined here as a mean absolute
deviation in $G$ band below $0.02$ \citep{maiz2023}. These two quantities
are field-level summary statistics and need not refer to exactly the
same stars. We split the sample by Galactic latitude $b$ into a
Galactic-plane bin $|b|<15^\circ$ and a high-latitude bin
$(|b|>30^\circ$, out of the plane), with $N_\star=100$ stars per Galactic-latitude bin
and typical binomial fraction uncertainties of order $\sim0.05$ at
$1\sigma$.}
{Lines of sight through the plane suffer more
reddening, {but we do not correct for this here.} We combine the two fractions into a
single Habitability Potential Index,}
\begin{equation}
{\rm HPI} = 0.5\,f_{\rm solar} + 0.5\,f_{\rm calm},
\label{eq:hpi}
\end{equation}
{where $f_{\rm solar}, f_{\rm calm} \in [0,1]$.
HPI also lies between 0 and 1, where values close to 0 indicate fields
with few solar-type or photometrically calm stars, while values close to
1 indicate fields rich in both types of favourable targets.
In the small Gaia DR3 samples used here (Table 
~\ref{tab:gaia_habitability_context}), the Galactic-plane bin has a
higher HPI  than the high-latitude bin, because it contains both a larger fraction of
late-G/K-colour stars and a larger fraction of photometrically calm
stars.  Under this simple two-feature ranking,
the Galactic-plane sample contains more targets with those  photometric
properties we regard as favourable for anomaly searches. However, the sample size is small and no reddening correction has been applied, so this comparison should be treated only as a proof-of-concept demonstration. The result should
therefore be read as ``where might LSST anomaly searches be easiest to
prioritise?'', not as ``where are habitable planets more common?''.}


\begin{table}[t]
\centering
\caption{Gaia DR3 sample properties.
{$N_\star$  is the number of stars per sample,
$f_{\rm solar}$ is the fraction  of 
late-G to K-type stars, 
$f_{\rm calm}$ is the fraction of stars with low
photometric variability}.
$\mathrm{HPI}$ is a normalised ranking variable indicating where anomaly searches may be easiest to prioritise.}
\label{tab:gaia_habitability_context}
\begin{tabular}{lcccc}
\toprule
{Sample} & $N_\star$ & $f_{\rm solar}$ & $f_{\rm calm}$ & {HPI} \\
\midrule
Galactic Plane        & 100 & $0.63\pm0.05$ & $0.80\pm0.04$ & 0.72 \\
High Galactic Latitude & 100 & $0.50\pm0.05$ & $0.68\pm0.05$ & 0.59 \\
\bottomrule
\end{tabular}
\end{table}

\section{Conclusion and outlook}
\label{sec:conclusion}

{We have presented three} prototype Rubin/LSST test cases --- KBO surface
colour changes, an injected VRE-like spectral feature, and multiband
time-domain brightness variability --- {showing that,} under idealised
simulated conditions, physically meaningful signals tend to
appear as a coordinated shift across several measurements at once,
rather than as an isolated odd measurement in a single band or
epoch. This work establishes the basic shape of that coordinated
signal and how to measure it; the critical next step is testing how
often realistic, ordinary astrophysical sources could accidentally
produce a similar-looking coordinated shift (i.e.\ characterising the
false-positive rate). Future work will: (i)~inject all three
diagnostics into realistic LSST observing cadences to quantify the
effect of irregular sampling and instrumental systematics;
(ii)~characterise false-positive rates against realistic populations
of ordinary astrophysical sources; (iii)~expand the population prior
to a full Rubin/LSST target list combining stellar colour,
photometric stability, and transiting-planet catalogues; and
(iv)~replace the idealised test features used here with more
physically motivated forward models.

\section*{Acknowledgments}

\noindent The Authors express gratitude to the Organizers of IAUS\,404
and the University of Kent for their support.
A.B.K.\ acknowledges funding provided by the University of
Belgrade -- Faculty of Mathematics (contract
451-03-33/2026-03/200104).
N.J.M.\ recognises the support of Europlanet 2024 RI, which has
received funding from the European Union's Horizon 2020 research and
innovation programme under grant agreement No.\,871149. A. {\' C}.: This work was produced by Fermi Forward Discovery Group, LLC under Contract No. 89243024CSC000002 with the U.S. Department of Energy, Office of Science, Office of High Energy Physics. The United States Government retains, and the publisher, by accepting the work for publication, acknowledges that the United States Government retains a non-exclusive, paid-up, irrevocable, world-wide license to publish or reproduce the published form of this work, or allow others to do so, for United States Government purposes. The Department of Energy will provide public access to these results of federally sponsored research in accordance with the DOE Public Access Plan (http://energy.gov/downloads/doe-public-access-plan).


\end{document}